# Growth and Characterization of Metalorganic Vapor-Phase Epitaxy-Grown β-(Al$_x$Ga$_{1-x}$)$_2$O$_3$/β-Ga$_2$O$_3$ Heterostructure Channels

Praneeth Ranga[1, a)], Arkka Bhattacharyya[1, a)], Adrian Chmielewski[2], Saurav Roy[1], Rujun Sun[1], Michael A. Scarpulla[1,3], Nasim Alem[2] and Sriram Krishnamoorthy[1]

[1]*Department of Electrical and Computer Engineering, The University of Utah, Salt Lake City, UT 84112, USA*

[2]*Department of Materials Science and Engineering, Pennsylvania State University, University Park, PA 16802, USA*

[3]*Department of Materials Science and Engineering, University of Utah, Salt Lake City, Utah 84112, USA*

[a)] *Praneeth Ranga and Arkka Bhattacharyya contributed equally*

E-mail: praneeth.ranga@utah.edu, sriram.krishnamoorthy@utah.edu

We report on the growth and characterization of metalorganic vapor-phase epitaxy-grown β-(Al$_x$Ga$_{1-x}$)$_2$O$_3$/β-Ga$_2$O$_3$ modulation-doped heterostructures. Electron channel is realized in the heterostructure by utilizing a delta-doped β-(Al$_x$Ga$_{1-x}$)$_2$O$_3$ barrier. Electron channel characteristics are studied using transfer length method, capacitance-voltage and Hall measurements. Hall sheet charge density of 1.06 x 10$^{13}$ cm$^{-2}$ and mobility of 111 cm$^2$/Vs is measured at room temperature. Fabricated transistor showed peak current of 22 mA/mm and on-off ratio of 8 x 10$^6$. Sheet resistance of 5.3 kΩ/Square is measured at room temperature, which includes contribution from a parallel channel in β-(Al$_x$Ga$_{1-x}$)$_2$O$_3$.

Ultrawide bandgap materials such as $\beta$-Ga$_2$O$_3$ has attracted a lot of interest because of their suitable properties for high-power electronics and deep-UV optoelectronic applications. The high bandgap (~4.6 eV) results in a very large predicted breakdown field strength of 6-8 MV/cm, which is much larger than other wide bandgap materials like GaN and SiC[1]. Significant advances have been made in growth[2–5], characterization, and fabrication of $\beta$-Ga$_2$O$_3$ devices within the last decade[6]. Vertical and lateral devices with high breakdown voltages and high critical fields have been demonstrated by multiple research groups[7–11].

Room temperature mobility of uniformly-doped $\beta$-Ga$_2$O$_3$ is limited by polar optical phonon scattering, which limits the maximum mobility to ~200 cm$^2$/Vs[12]. In modulation-doped 2DEG channel, mobility is not limited by impurity scattering unlike doped semiconductors. This is due to the absence of ionized impurity donors in the electron channel. For realizing a 2DEG, growth of high-quality modulation-doped $\beta$-(Al$_x$Ga$_{1-x}$)$_2$O$_3$/$\beta$-Ga$_2$O$_3$ heterostructure with sharp dopant profile is necessary. Theoretical studies indicate that $\beta$-(Al$_x$Ga$_{1-x}$)$_2$O$_3$ is stable up to a composition of ~x = 0.8 which has a bandgap of ~6.5 eV[13]. Moreover, DFT calculations indicate that n-type shallow doping is achievable for the entire composition range of stable $\beta$-(Al$_x$Ga$_{1-x}$)$_2$O$_3$[14]. All the above material properties suggest that formation of 2DEG at $\beta$-(Al$_x$Ga$_{1-x}$)$_2$O$_3$/$\beta$-Ga$_2$O$_3$ is very favorable. Recently reported transport calculations performed by Kumar et.al indicate that mobility of a 2DEG can significantly exceed that of bulk $\beta$-Ga$_2$O$_3$[15]. This is expected to happen when the 2DEG sheet charge exceeds 5 x 10$^{12}$ cm$^{-2}$. At high charge densities (n$_s$ > 5 x 10$^{12}$ cm$^{-2}$ ), the plasmon screening of LO (longitudinal optical) phonons leads to increase in polar optical phonon (POP) limited mobility[15]. Having a high 2DEG sheet charge and mobility can lead to a significant improvement in device performance over conventional $\beta$-Ga$_2$O$_3$ devices.

Demonstration of $\beta$-(Al$_x$Ga$_{1-x}$)$_2$O$_3$/$\beta$-Ga$_2$O$_3$ heterostructure modulation-doped field effect transistors (MODFET) has already been made using MBE-grown material[16,17]. All the current literature is based on MBE-grown $\beta$-(Al$_x$Ga$_{1-x}$)$_2$O$_3$/$\beta$-Ga$_2$O$_3$ heterostructures[16–21]. 2DEG sheet charge densities between 1x10$^{12}$ - 5 x10$^{12}$ cm$^{-2}$ and mobilities of 75 -180 cm$^2$/Vs have been achieved using MBE technique[16,18–21]. Currently the maximum sheet charge density reported in MBE-grown β-$(Al_xGa_{1-x})_2O_3$/β-$Ga_2O_3$ is less than 5 x $10^{12}$ cm$^{-2}$ for a single heterostructure without parallel channel in β-$(Al_xGa_{1-x})_2O_3$[20]. Recently, MOVPE (metalorganic vapor-phase epitaxy) has emerged as a promising technique for high-quality β-$Ga_2O_3$. Uniformly-doped β-$Ga_2O_3$ films with high room temperature mobility values have been demonstrated[3,22–24]. High composition MOVPE-grown (100)-oriented β-$(Al_xGa_{1-x})_2O_3$ films with x ~ 0.52 has been realized recently[25]. N-type doping of MOVPE-grown β-$(Al_xGa_{1-x})_2O_3$ films with composition up to x ~ 0.3 has already been demonstrated[26,27]. In addition, delta-doped β-$Ga_2O_3$ films with sheet charge up to 1 x $10^{13}$ cm$^{-2}$ is reported using MOVPE[28]. For achieving sharp dopant profiles with low FWHM (full width at half maximum), it is necessary to suppress surface segregation of donor atoms. Our recent work on MOVPE-grown delta-doped films grown at low temperatures suggests that by suppressing surface segregation of Si delta sheets with FWHM comparable to MBE-grown films can be realized[29]. All the above factors indicate the promise of MOVPE technique for studying high charge density β-$(Al_xGa_{1-x})_2O_3$/β-$Ga_2O_3$ heterostructures.

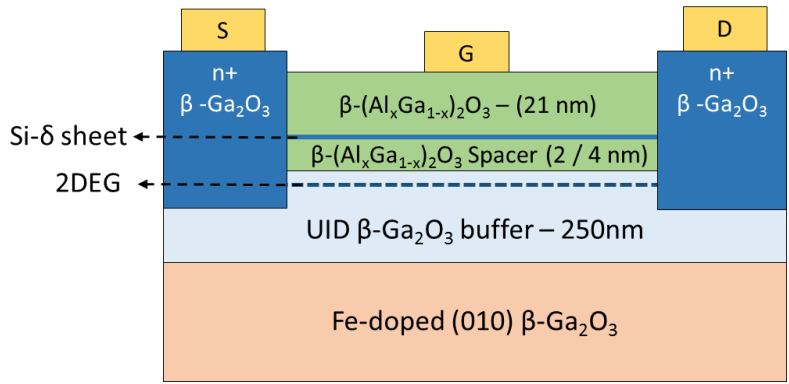

**Fig.1** Schematic of MOVPE-grown β-$(Al_xGa_{1-x})_2O_3$/ β-$Ga_2O_3$ heterostructure FET (HFET) with regrown ohmic contacts. (Sample A spacer layer thickness- 4 nm; Sample B spacer layer thickness- 2 nm).

β-$(Al_xGa_{1-x})_2O_3$/β-$Ga_2O_3$ heterostructures are grown using Agnitron Agilis using TEGa (Triethyl Gallium), TMAl (Trimethyl Aluminium) and $O_2$(oxygen) as precursors and Argon as carrier gas. The following growth parameters are utilized –Alkyl flow – 5.26 μmol/min, $O_2$ – 500 sccm, Pressure - 15 Torr, Temperature - 650 °C. Growth is performed on Fe-doped (010) β-$Ga_2O_3$ substrates from Novel Crystal Technology. As-received substrates are cleaned with Acetone, Methanol and DI water followed a HF dip for 20 mins. A schematic of the grown β-$(Al_xGa_{1-x})_2O_3$/β-$Ga_2O_3$ heterostructure is shown in fig.1. The film consists of a 250 nm UID β-$Ga_2O_3$ buffer layer followed by a thin β-$(Al_xGa_{1-x})_2O_3$ spacer layer (2 / 4 nm) and a thick β-$(Al_xGa_{1-x})_2O_3$ barrier (21 nm). Next, delta doping of β-$(Al_xGa_{1-x})_2O_3$ layer is performed by using a growth interruption process. The process consists of a growth interruption step with a pre- and post-purge steps (30 secs) before and after the delta doping. The delta sheet density is controlled by changing the silane flow period (60 secs) and the silane gas flow (A - 26 nmol/min, B – 34.7 nmol/min). Additional details and study of the delta doping process in β-$Ga_2O_3$ are reported elsewhere[28,29]. After the delta doping process, growth of additional 21 nm β-$(Al_xGa_{1-x})_2O_3$ barrier is continued until the desired total thickness is reached (~ 23 - 25 nm including the spacer layer). The composition of the β-$(Al_xGa_{1-x})_2O_3$ barrier is controlled by setting the [Al]/([Ga] + [Al]) molar ratio to x~ 18 % (sample A) and x ~ 25 % (sample B). In sample B, the β-$(Al_xGa_{1-x})_2O_3$ molar ratio is increased to 25% and silane flow is increased by 25 % while reducing the spacer thickness to 2 nm. Composition of the β-$(Al_xGa_{1-x})_2O_3$ barrier is verified using XRD (X-ray diffraction)[30]. The measured composition is x-0.20 for sample A and x-0.27 for sample B, which is close to the precursor molar ratio (see supplementary).

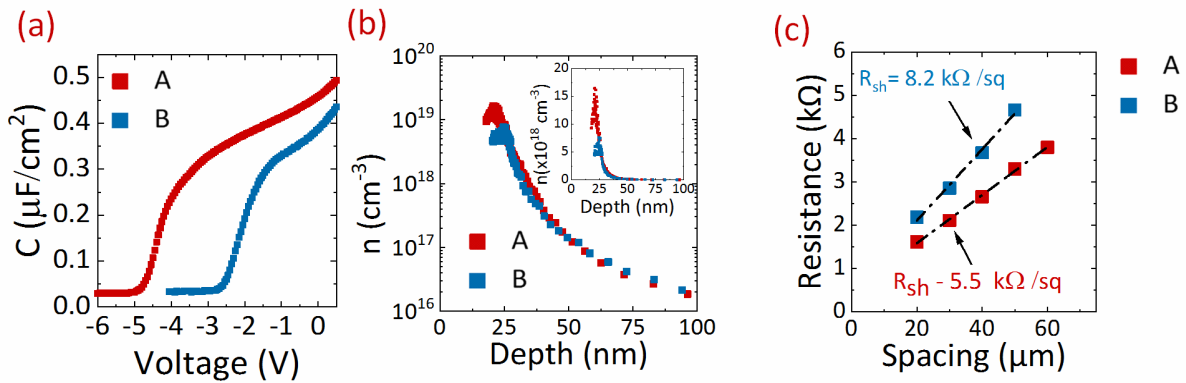

**Fig.2** (a) Capacitance- Voltage measurement at room temperature (b) Extracted apparent charge density profiles for samples A and B (inset- linear scale) (c) TLM measurements of $\beta$-$(Al_xGa_{1-x})_2O_3/\beta$-$Ga_2O_3$ heterostructures.

FET, TLM and Van der Pauw structures were realized after mesa isolation using dry etching with $SF_6/Ar$ chemistry (35 sccm/5 sccm, 150W RF and 600W ICP powers). Next the samples are prepared for $n^+$ layer regrowth for ohmic contact formation. The samples are patterned using $Ni/SiO_2$ mask followed by a 50 nm etch using $SF_6/Ar$ plasma (35 sccm/5 sccm, 150 W RF power). After the etch, Ni mask layer is removed using wet etch (diluted aqua regia) and samples are loaded in to the MOVPE reactor. Low temperature $n^+$ regrowth is performed using MOVPE at a growth temperature of 600 °C. The low growth temperature is chosen to minimize any potential damage to the epitaxial layer. The doping of the $n^+$ layer is set to 8 x $10^{19}$ $cm^{-3}$ and the thickness of the $n^+$ layer is 150 nm. After the completion of the $n^+$ regrowth process, the $SiO_2$ hard mask is removed by a HF dip. Standard photolithography process is utilized for patterning Ohmic source/drain and Schottky gate pads. After the deposition of Ohmic contacts, the contacts are annealed using a RTA (Rapid thermal anneal - 90s, 450 °C) system under nitrogen environment. Both the Ohmic (30 nm Ti /100 nm Au /30 nm Ni) and Schottky contacts (30 nm Ni/50 nm Au/30 nm Ni) are deposited by ebeam evaporation.

CV, TLM and Hall measurements are utilized to independently measure apparent charge density profile, sheet charge, mobility, and sheet resistance. CV measurements performed on samples A and B are shown in fig.2(a). The apparent charge density extracted from the charge profile (including forward tail of the apparent charge profile measured until + 0.5 V) decreased from 1.1 x $10^{13}$ $cm^{-2}$ to 5.7 x $10^{12}$ $cm^{-2}$ between sample A and B. Extracted apparent charge density profile of the 2DEG sheet charge is plotted in fig.2(b). Room temperature Hall measurements are also performed to measure sheet charge and mobility. Details of all the electrical measurements are summarized in Table. 1. Hall sheet charges of 1.06 x $10^{13}$ $cm^{-2}$ and 6.4 x $10^{12}$ $cm^{-2}$ are recorded for sample A and B. Room temperature electron mobility values of 111 $cm^2/Vs$ and 125 $cm^2/Vs$ are measured for samples A and B, respectively. Additionally, TLM measurements are performed to extract electron channel sheet resistance. TLM sheet resistance values of 5.5 kΩ/Square and 8.2 kΩ/Square are measured for samples A and B. The sheet charge, sheet resistance and channel mobility of Hall and TLM measurements   are listed in Table 1. These electrical measurements indicate that all the measured parameters can be directly attributed to the β-$(Al_xGa_{1-x})_2O_3$/β-$Ga_2O_3$ heterostucture channel (2DEG and parallel channel in the alloy barrier).

**Table I.    Hall, CV and TLM characterization of β-$(Al_xGa_{1-x})_2O_3$/β-$Ga_2O_3$ heterostucture channel**

| Sample | Spacer thickness (nm) | Hall sheet charge (x $10^{12}$ cm$^{-2}$) | | Hall mobility cm$^2$/Vs | | RT Hall Sheet resistance (kΩ/square) | RT TLM Sheet resistance (kΩ/square) |
|--------|------|--------|------|-------|------|------|------|
|        |      | 300 K  | 77 K | 300K  | 77 K |      |      |
| A      | 4    | 10.6   | 3    | 111   | 1680 | 5.3  | 5.5  |
| B      | 2    | 6.4    | 3.1  | 125   | 689  | 7.7  | 8.2  |

To understand the nature of the heterostructure electron channel, low temperature Hall measurements are performed at liquid nitrogen temperatures (77K). For an ideal 2DEG with no parallel channel in the β-$(Al_xGa_{1-x})_2O_3$ layer, the Hall measured sheet charge density is not expected to freeze out upon reaching cryogenic temperatures[18]. The sheet charge of samples A and B reduced to ~ 3 x $10^{12}$ cm$^{-2}$ at 77 K. Correspondingly, the Hall mobility increased to 1680 cm$^2$/Vs and 689 cm$^2$/Vs for samples A and B.    The Hall measured sheet charge density reduced to 1.2 kΩ/square(sample A) and 2.9 kΩ/square(sample B) at 77K. This indicates that there is a significant amount of donor freezeout at 77 K. This reduction in sheet charge is attributed to freezeout of parallel channel in β-$(Al_xGa_{1-x})_2O_3$. The high low-temperature mobility observed in sample A could be potentially attributed to improved material quality of low composition β-$(Al_{0.2}Ga_{0.8})_2O_3$/β-$Ga_2O_3$ heterointerface, compared to the higher alloy composition in sample B.    Additional hall measurements through out the entire temperature range may give us more insight on carrier freeze out in β-$(Al_xGa_{1-x})_2O_3$ layer. In a delta-doped heterostructure with a narrow dopant profile, all the donor atoms are located in the β-$(Al_xGa_{1-x})_2O_3$ barrier layer. However, attaining silicon delta sheets with low

FWHM is challenging[28]. Because of the spread-out dopant profile, non-negligible amount of charge may end up in the UID β-Ga$_2$O$_3$ layer. A finite amount of charge reduction could result from carrier freeze out in the UID β-Ga$_2$O$_3$ layer. Reports of charge freeze out at low temperature have been observed in β-(Al$_x$Ga$_{1-x}$)$_2$O$_3$/β-Ga$_2$O$_3$ single and double heterostructures[18,19,21] and uniformly doped β-(Al$_x$Ga$_{1-x}$)$_2$O$_3$ thin films[27]. Based on theory, it is expected that formation of parasitic channel becomes more favorable with increasing spacer thickness. Measured Hall data also indicates similar trend. However, modeling the electron charge in this case is quite challenging due to the combination of several factors such as non-ideal delta sheet profile with spread out Si donors in the alloy barrier, incomplete ionization in the alloy barrier and unintentional doping of the UID channel due to spread of the donors. Also, the donor ionization is expected to change with different Al composition. Additionally, STEM investigations revealed formation of defects in the β-(Al$_{0.28}$Ga$_{0.72}$)$_2$O$_3$ layer > 5nm away from the heterointerface (see supplementary). The above analysis indicates that getting a very high 2DEG charge in β-(Al$_x$Ga$_{1-x}$)$_2$O$_3$/β-Ga$_2$O$_3$ heterostructures is still challenging. For attaining high charge densities, it is important to realize a sharp dopant profiles in β-(Al$_x$Ga$_{1-x}$)$_2$O$_3$ barrier with a high conduction band offset at the -(Al$_x$Ga$_{1-x}$)$_2$O$_3$/ Ga$_2$O$_3$ heterojunction . Higher Al composition barrier in conjunction with thin spacer layer and sharp doping profile would be necessary to leverage the predicted high electron mobilities in β-Ga$_2$O$_3$ 2DEGs.

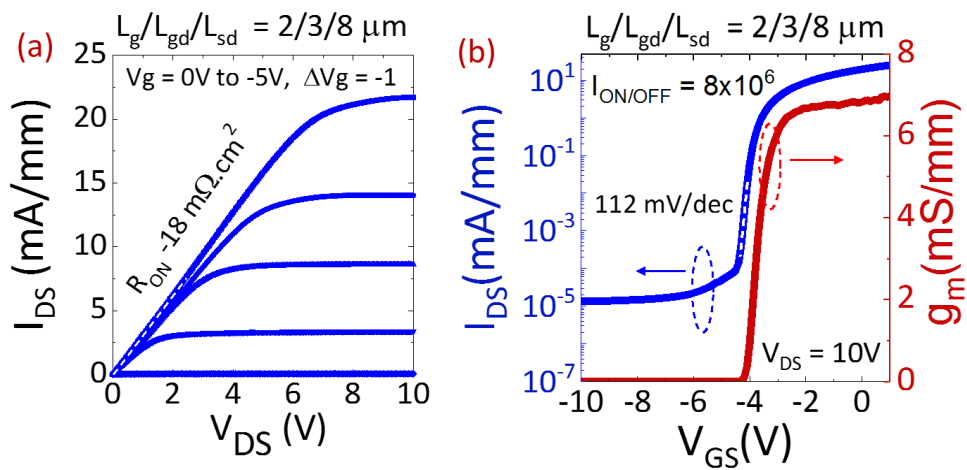

**Fig.3** (a) Output and (b) transfer characteristics of β-(Al$_x$Ga$_{1-x}$)$_2$O$_3$/ β-Ga$_2$O$_3$ heterostructure FET showing peak current density of 22 mA/mm (V$_{ds}$ = 10 V, V$_g$ = 0 V ) and transconductance of 7 mS/mm (V$_{ds}$ = 10 V, V$_g$ = 0 V )

Lateral heterojunction field effect transistors (HFETs) are fabricated with n+ regrown contacts (sample A) with L$_g$, L$_{sd}$, L$_{gd}$ of 2 μm ,8 μm and 3 μm respectively. Output and transfer characteristics of the FET are shown in Fig. 3(a) and 3(b). The devices showed a peak current of 22 mA/mm at zero gate bias and drain bias of 10 V. From the TLM measurements, the contact resistance to the heterostructure channel was extracted to be as high as 16.5 ohm.mm. MOVPE films grown at low temperature showed high mobility and low compensation[24]. TLM measurements on regrown n+ layer without any channel layer, showed a ρ$_c$ ~ 10$^{-6}$ Ω –cm$^2$   indicating low resistance n+ Ga$_2$O$_3$ /Ti contact. The high R$_c$ is attributed to etch damage and charge depletion due to Fluorine ions[31] resulting from the SF$_6$/Ar dry etch step used before the contact regrowth step. Switching to BCl$_3$/Ar chemistry-based dry etch step could lead to lower contact resistivity. The peak current is currently limited by the device dimension and the performance of the source/drain ohmic contacts. The device showed good pinch off characteristics with a pinch off voltage close of -4 V, correlating well with the CV measurements. The device also showed a high on-off ratio of 8 x 10$^6$ and sub-threshold swing of 112 mV/dec. A peak transconductance of 7 mS/mm is measured at zero gate bias.

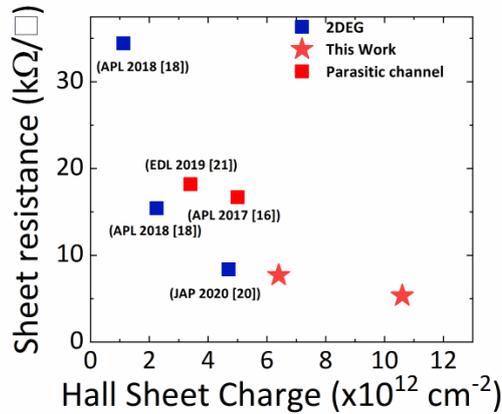

**Fig. 4** Comparison of measured Hall sheet resistance as a function of charge density for a single β-$(Al_xGa_{1-x})_2O_3$/β-$Ga_2O_3$ heterostructure channel.

The measured room temperature sheet resistance values are plotted as a function of the charge density in fig.4 (see supplementary for Hall measurement data). In this work, we report a low sheet resistance value of 5.3 kΩ/Square for a single β-$(Al_xGa_{1-x})_2O_3$/ β-$Ga_2O_3$ heterostructure. Additionally, we report β-$(Al_xGa_{1-x})_2O_3$/β-$Ga_2O_3$ heterostructure device grown and fabricated completely based on MOVPE process, including contact regrowth. However, the complete charge cannot be ascribed to modulation-doped carriers because of parallel channel in β-$(Al_xGa_{1-x})_2O_3$ layer as evidenced from observed carrier freezeout at low temperature. Nevertheless, the low sheet resistance value obtained in this work, with room temperature mobilites exceeding 100 cm$^2$/Vs is a promising step towards high performance MOVPE-grown β-$(Al_xGa_{1-x})_2O_3$/β-$Ga_2O_3$ modulation-doped devices.

In conclusion, we report on growth and characterization of MOVPE-grown β-$(Al_xGa_{1-x})_2O_3$/ β-$Ga_2O_3$ heterostructure channel with low sheet resistance. Electrical characteristics of the heterostucture channel are measured using TLM, CV and Hall measurements. Room temperature Hall measurements showed a high sheet charge of 6.4 x10$^{12}$ – 1.06 x 10$^{13}$ cm$^{-2}$ and mobility of 111-125 cm$^2$/Vs. STEM investigation of β-$(Al_xGa_{1-x})_2O_3$/β-$Ga_2O_3$ heterostructure showed formation of defects away from the interface. FET showed a peak current density of 22 mA/mm and on-off ratio of 8x10$^6$.

**Acknowledgments**

This material is based upon work supported by the Air Force Office of Scientific Research under award number FA9550-18-1-0507 and monitored by Dr. Ali Sayir. Any opinions, findings, conclusions, or recommendations expressed in this material are those of the authors and do not necessarily reflect the views of the United States Air Force. Praneeth Ranga acknowledges support from University of Utah Graduate Research Fellowship 2020-2021. This work was performed in part at the Utah Nanofab sponsored by the College of Engineering and the Office of the Vice President for Research. The authors thank the Air Force Research Laboratory's Sensors Directorate for their discussions with them. The electron microscopy work was performed in the Materials Characterization lab (MCL) at the Materials Research Institute (MRI) at the Pennsylvania State University. The work at PSU was supported by the AFOSR program FA9550-18-1-0277 (GAME MURI, Dr. Ali Sayir, Program Manager).

## Figure Captions

**Fig. 1.** Schematic of MOVPE-grown $\beta$-(Al$_x$Ga$_{1-x}$)$_2$O$_3$/ $\beta$-Ga$_2$O$_3$ heterostructure FET (HFET) with regrown ohmic contacts. (Sample A spacer layer thickness- 4 nm; Sample B spacer layer thickness- 2 nm).

**Fig. 2.** (a) Capacitance- Voltage measurements at room temperature (b) Extracted apparent charge density profiles for samples A and B (inset-linear scale)(c) TLM measurements of $\beta$-(Al$_x$Ga$_{1-x}$)$_2$O$_3$/$\beta$-Ga$_2$O$_3$ heterostructures.

**Fig. 3.** (a) Output and (b) transfer characteristics of $\beta$-(Al$_x$Ga$_{1-x}$)$_2$O$_3$/ $\beta$-Ga$_2$O$_3$ heterostructure FET showing peak current density of 22 mA/mm (V$_{ds}$ = 10 V, V$_g$ = 0 V) and transconductance of 7 mS/mm (V$_{ds}$ = 10 V, V$_g$ = 0 V )

**Fig. 4.** Comparison of measured Hall sheet resistance as a function of charge density for a single $\beta$-(Al$_x$Ga$_{1-x}$)$_2$O$_3$/$\beta$-Ga$_2$O$_3$ heterostructure channel.

**Figure 1**

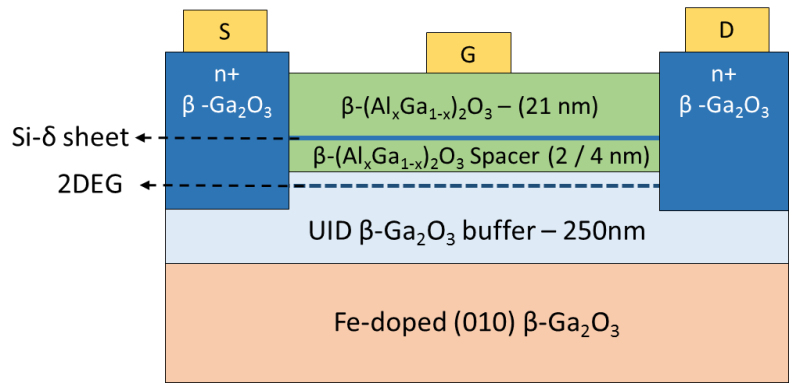

**Figure 2**

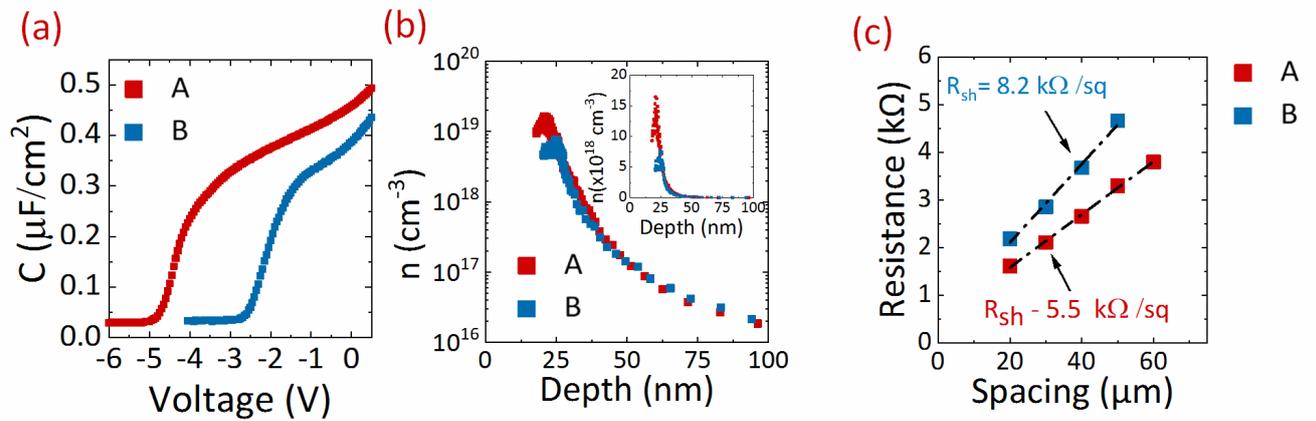

**Figure 3**

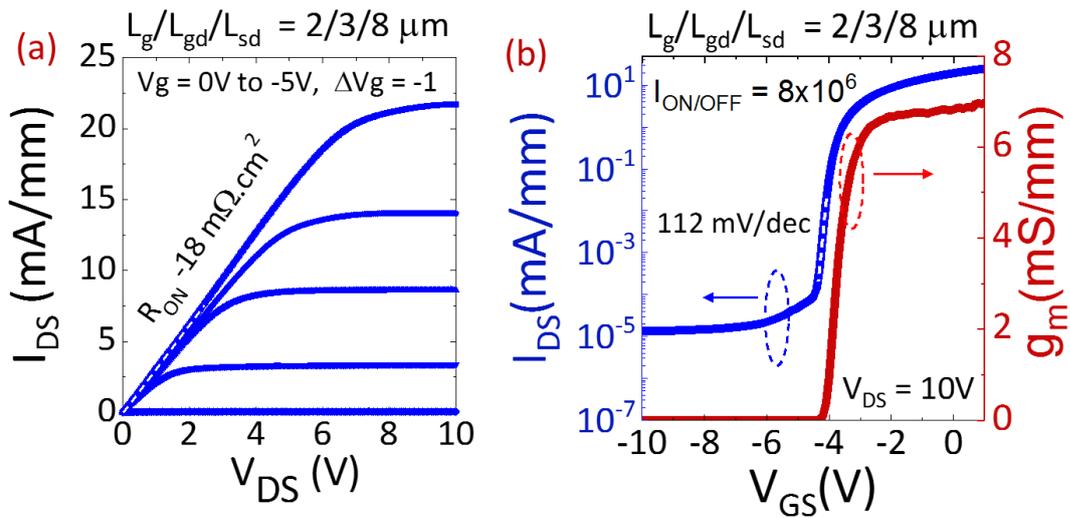

**Figure 4**

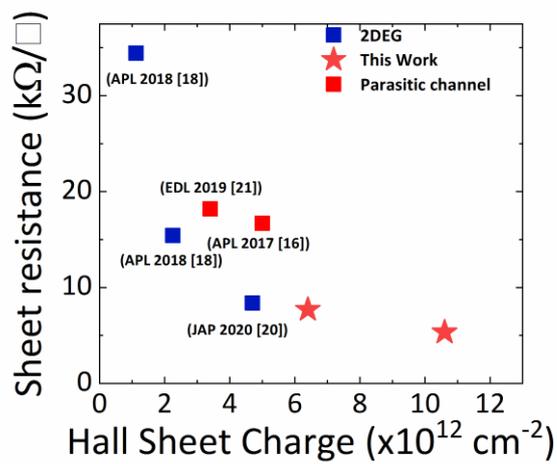

# Growth and Characterization of Metalorganic Vapor-Phase Epitaxy-Grown β-(AlxGa1-x)2O3/β-Ga2O3 Heterostructure Channels

Praneeth Ranga[1, a)], Arkka Bhattacharyya[1, a)], Adrian Chmielewski[2], Saurav Roy[1], Rujun Sun[1], Michael A. Scarpulla[1,3], Nasim Alem[2] and Sriram Krishnamoorthy[1]

[1] *Department of Electrical and Computer Engineering, The University of Utah, Salt Lake City, UT 84112, United States of America*

[2]*Department of Materials Science and Engineering, Pennsylvania State University, University Park, State College, PA 16802, USA*

[3] *Department of Materials Science and Engineering, University of Utah, Salt Lake City, Utah 84112, USA*

a)    *Praneeth Ranga and   Arkka Bhattacharyya contributed equally to this work.*

## Supplementary information

**X ray diffraction scan of β-(AlxGa1-x)2O3/Ga2O3 heterostructures**

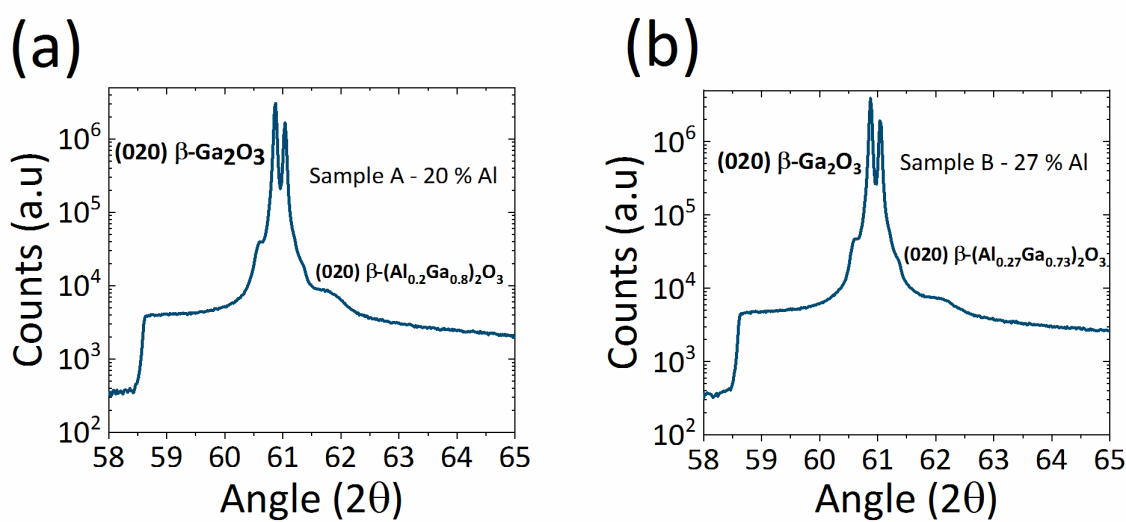

Fig.S1 XRD scan of β-(AlxGa1-x)2O3/β-Ga2O3 heterostructure (a) Sample A (20% - Al) (b) Sample B (27 % - Al). Incident X-ray has both Kα1 and K α2 wavelengths.

**STEM images of β-(Al₀.₂₈Ga₀.₇₂)₂O₃/β-Ga₂O₃  interface**

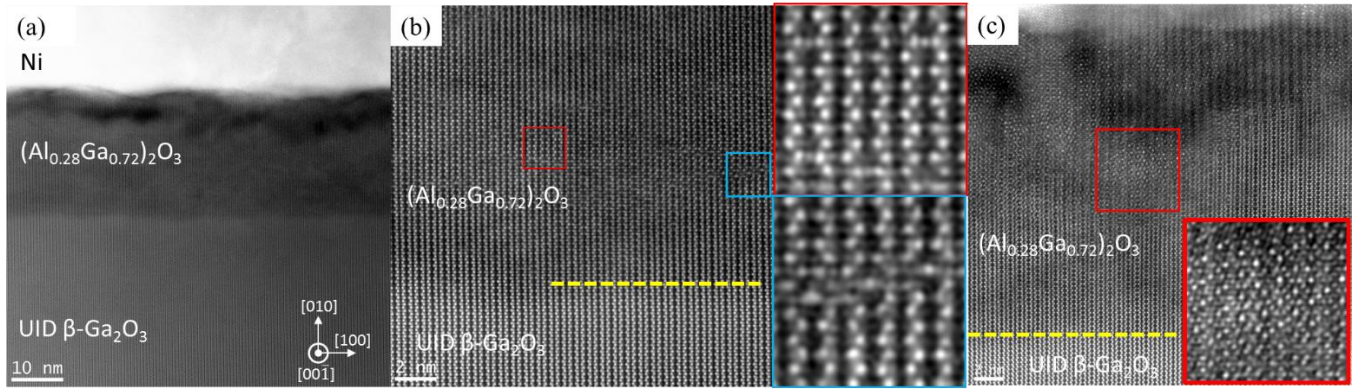

Fig. S2 (a) HAADF-STEM image of the β-(Al$_{0.28}$Ga$_{0.72}$)$_2$O$_3$/β-Ga$_2$O$_3$  interface in the [00$\bar{1}$] projection at low magnification (b) zoomed-in image of the    β-(Al$_{0.28}$Ga$_{0.72}$)$_2$O$_3$/β-Ga$_2$O$_3$ heterostructure showing defect formation in    β-(Al$_{0.28}$Ga$_{0.72}$)$_2$O$_3$ layer (inset: high magnification image of the defect region) (c) HAADF-STEM image of    β-(Al$_{0.28}$Ga$_{0.72}$)$_2$O$_3$/β-Ga$_2$O$_3$    heterostructure showing defect formation (inset: high magnification image of the defective region)

High resolution high angle annular dark field-scanning transmission electron microscopy (HAADF- STEM) investigations are performed on β-(Al$_{0.28}$Ga$_{0.72}$)$_2$O$_3$/β-Ga$_2$O$_3$ grown on Sn-doped (010) β-Ga$_2$O$_3$  structures with growth conditions identical to sample B (2 nm spacer). HAADF-STEM imaging is carried out using a FEI Titan G2 60-300 transmission electron microscope (TEM) at 300kV. A condenser aperture of 70um is used with a convergence angle of 30 mrads and the annular detector collection angles in the 42-250 mrad range. The camera length is set to 1.15m and the probe current is approximately 80 pA. As the contrast is proportional to the Z number of the atom in the STEM mode, the heavier the atom, the brighter it will appear on the projected image. Figure S2(a) is a HAADF-STEM image of the β-(Al$_{0.28}$Ga$_{0.72}$)$_2$O$_3$/β-Ga$_2$O$_3$   interface in the [00$\bar{1}$] projection at low magnification. This zone axis allows to have the distance between Ga atoms easily resolved by the TEM. The top of the image corresponds to the Nickel contact layer whereas the β-$(Al_{0.28}Ga_{0.72})_2O_3$/β-$Ga_2O_3$ interface is easily observable due to the contrast change. A zoom-in image of the interface is shown in fig. S2(b). The yellow dashed lines represents the heterostructure interface. The β-$Ga_2O_3$ substrate has a homogeneous bright contrast whereas the β-$(Al_{0.28}Ga_{0.72})_2O_3$ film is darker overall which is due to the presence of lighter Al atoms. Interestingly, different types of point defects are observed in the β-$(Al_{0.28}Ga_{0.72})_2O_3$ film such as Al/Ga interstitials and Ga vacancies. Some of these defects have already been studied in the literature. For instance, Johnson et al. have shown the formation of Ga interstitial sitting in between two Ga vacancies creating a 2VGa-Ga$_i$ complex in Sn doped β-$Ga_2O_3$ [1]. More recently, A. Chmielewski et al. reported the formation of di-interstitial di-vacancy complexes at the MBE grown β-$(Al_{0.2}Ga_{0.8})_2O_3$/β-$Ga_2O_3$ interface that are created by either an Al or Ga interstitial in the tetrahedral site[2].

Figure. S2(c) is a HAADF-STEM image of another area of the sample. A zoom-in image of the defective area is shown in the red square in which a similar structure to the $\gamma$–phase is observed. A mixture of β and γ phases in MOCVD grown β-$(Al_xGa_{1-x})_2O_3$, when Al composition ranged between 27% and 40%, has already been reported in the literature [3]. Although these contrast are very similar to those of a $\gamma$–phase, similar structures were observed when two β-lattices with a relative shift of 2.72 Å in the ⟨102⟩ direction are superimposed along the ⟨010⟩ viewing direction[4]. As of now, there is no general consensus towards the understanding of this contrast and a deeper analysis will be necessary to have a better comprehension of it. However, here, it is clearly observable that these defects are present further from the β-$(Al_{0.28}Ga_{0.72})_2O_3$/β-$Ga_2O_3$ interface as shown by the inserts in Figure S2(b).

**Low temperature capacitance-voltage extracted apparent charge density**

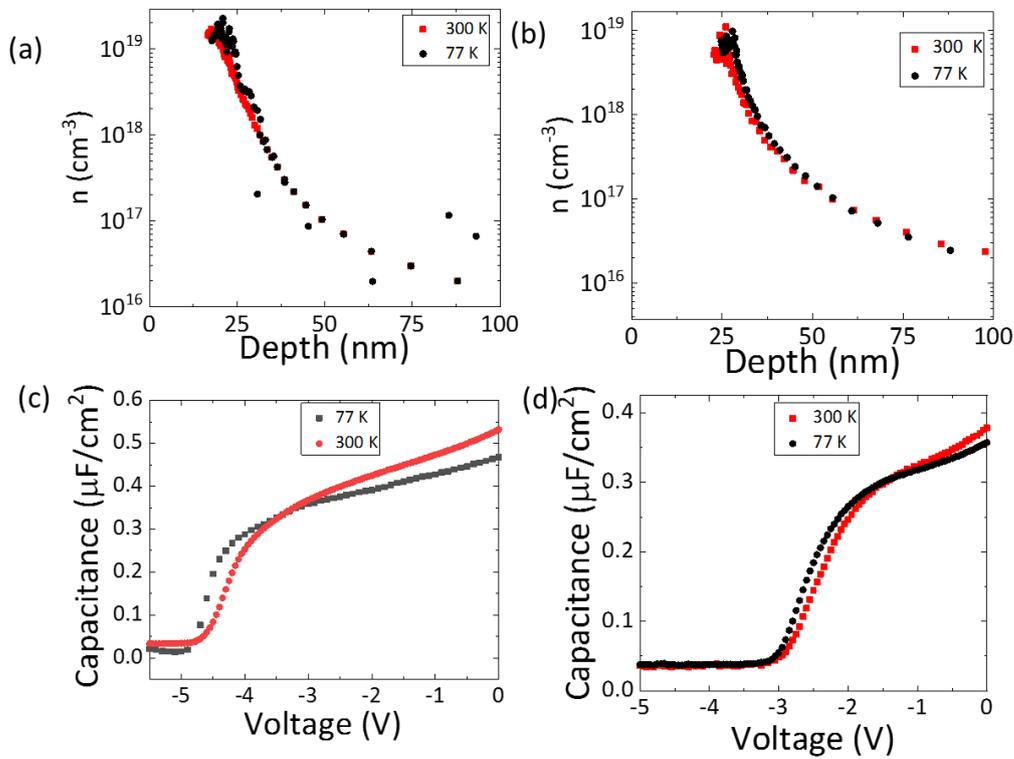

Fig.S3 Capacitance-voltage data of a) sample A and b) Sample B measured at 300 K and 77 K. Apparent charge density profile extracted from capacitance-voltage measurements at 77 K and 300 K (c) Sample A (d) Sample B

CV measurements of the β-(Al$_x$Ga$_{1-x}$)$_2$O$_3$/β-Ga$_2$O$_3$ heterostructure revealed no significant difference upon going from 300 K to 77 K. However, low temperature hall measurements clearly showed carrier freeze out. This discrepancy can be explained by understanding that CV extracted charge density doesn't follow the actual carrier profile at 77 K. Instead, CV measures the apparent charge profile (donor density), leading to observation of high apparent charge density at 77 K, whereas, Hall measurement can only measure free carriers present in the β-(Al$_x$Ga$_{1-x}$)$_2$O$_3$/β-Ga$_2$O$_3$ channel layer. It should be noted that the amount of ionization of the donors also change with reverse bias, further complicating the interpretation of CV measurements.

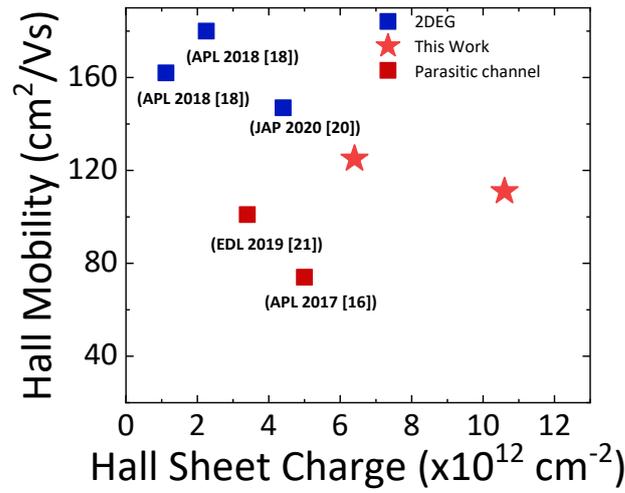

Fig. S4 Hall mobility plotted against Hall sheet charge for β-(Al$_x$Ga$_{1-x}$)$_2$O$_3$/β-Ga$_2$O$_3$ heterostructure in literature